\documentclass[preprint,aps]{revtex4}
\begin{document}
\title{The Production and Decay of Heavy Dimesons}
\author{Mitja Rosina}
\affiliation{Faculty of Mathematics and Physics, University of Ljubljana,\\
     Jadranska 19, P.O.~Box 2964, 1001 Ljubljana, Slovenia\\
Jo\v{z}ef~Stefan Institute, 1000 Ljubljana, Slovenia}
\author{Damijan Janc}
\affiliation{Jo\v{z}ef~Stefan Institute, 
              1000 Ljubljana, Slovenia}
\author{Daniele Treleani and Alessio~Del~Fabbro}
\affiliation{Dipartimento di Fisica Teorica Universit\`a di Trieste,\\
     Strada Costiera 11, Miramare-Grignano \\ and INFN, Sezione di Trieste, I-34014 Trieste, Italy\\}
\begin{abstract}
          We show that it is necessary to go beyond a single hadron
          (beyond the quark-antiquark or three-quark systems) 
          in order to distinguish the colour structure of the
          effective quark-quark interaction and the relevance 
          of 3-body forces. We critically discuss the proposed
          models which suggest the dimeson ${\rm bb\bar{u}\bar{d}}$
          to be bound by $\sim 100$ MeV and the
          ${\rm cc\bar{u}\bar{d}}$ dimeson to be unbound.
          Only experiment can judge. We estimate the probability
          of producing ${\rm bb\bar{u}\bar{d}}$ at LHC by double
          gluon-gluon fusion and search for a characteristic decay.
\end{abstract}

\maketitle


\section{MOTIVATION}
There is a strong motivation to understand the effective interaction 
between heavy quarks (and antiquarks) since it is expected to be 
``cleaner'' than between light quarks. For heavy particles the nonrelativistic
constituent quark model is more acceptable, the perturbative QCD contributions
(such as one-gluon-exchange) is more adequate and chiral fields are less
important.

While the effective interaction between a heavy quark and a heavy antiquark
has been reasonably well studied and fitted by the charmonium and bottomium
spectra, there is no free diquark to study the effective interaction
between two heavy quarks. One has to dress the diquark in order to
obtain a colour singlet object. Therefore, the double-heavy baryons
ccq, bcq, and bbq (q=u,d, or s), as well as the double-heavy dimesons 
cc$\bar{{\rm q}}\bar{{\rm q}}$, bc$\bar{{\rm q}}\bar{{\rm q}}$
and bb$\bar{{\rm q}}\bar{{\rm q}}$ (also called tetraquarks) 
can be considered as a laboratory to study the properties of the diquark.
They have not yet been seen experimentally (except some tentative
signals \cite{Selex}). The purpose of our study
is to demonstrate, how important information for  quark models they
would offer, and to stimulate the experimentalists to invest due
efforts to discover them in near future .

It is straightforward to extrapolate the one-gluon-exchange (OGE) interaction
from Q$\bar{{\rm Q}}$ to QQ (Q= any quark). The charge conjugation
changes the $\bar{{\rm Q}}$ antitriplet to Q triplet. Then the colour
factor $\lambda\cdot\lambda/4 = -4/3$ for the Q$\bar{{\rm Q}}$
singlet changes to $-2/3$ for the QQ antitriplet 
(the ``$V_{{\rm QQ}}={1\over2}V_{{\rm Q}\bar{{\rm Q}}}$ {\it rule''}).

On the other hand, it is questionable whether the (linear) confining
potential should also possess such a colour factor and obey the
$V_{{\rm QQ}}={1\over2}V_{{\rm Q}\bar{{\rm Q}}}$ rule. The fact that
the ground state energies and some excited states of light and heavy
baryons are reasonably well reproduced with such a ``universal''
OGE + confining effective interaction is encouraging \cite{SB} 
but not conclusive. There may be other mechanisms for the 
$V_{{\rm QQ}}={1\over2}V_{{\rm Q}\bar{{\rm Q}}}$ rule. For example,
the flux tubes in a ``Mercedes" configuration can be mimicked by
twice weaker two-body flux lines since the length of the arms of the
``Mercedes'' is approximately half the length of the circumference
of the triangle. The colour singlet 3-quark system is insensitive to the 
features of the $colour\cdot colour$ operator since it is just 
a constant in the 3-body singlet representation. To explore the
colour structure of the effective interaction one has to go beyond
mesons and baryons to dimesons and other exotics.

Moreover, there may be other contributions to the effective interaction.
Regarding the short-range part, the important effect is the strong
spin-spin splitting which can be due to OGE, pion exchange or some instanton
effects, the relative importance of which is not yet clear.
Light baryon spectra are better reproduced by the  meson
exchange interaction \cite{Graz} than by the OGE interaction
but the extension to the heavy baryons and to mesons is still uncertain. 

The study of double-heavy baryons and of double-heavy dimesons are
complementary. They both refer to the colour triplet heavy diquark 
and try to determine the strength of the interaction and test 
the ``$V_{{\rm QQ}}={1\over2}V_{{\rm Q}\bar{{\rm Q}}}$ rule''. The dimeson,
however, possesses also a ($6,\bar{6}$) configuration with a sextet
heavy diquark and antisextet cloud of two light antiquarks; 
configuration mixing offers the opportunity to test the
colour structure of the interaction. 

Furthermore, there are theoretical reasons for three-body and four-body
forces. Three-body forces have been shown to be welcome to improve 
the absolute position of baryons with respect to mesons \cite{SB}.
The dimeson can give a better hint about the relative strength of the 
three-body interaction
since there are 4 three-body contributions compared to just one in a baryon.

In next Section we give arguments why we expect the
bb$\bar{{\rm u}}\bar{{\rm d}}$ dimeson to be strongly bound
while the cc$\bar{{\rm u}}\bar{{\rm d}}$, bc$\bar{{\rm u}}\bar{{\rm d}}$
and others are most likely unbound. In the third Section we estimate
the chances to produce the bb$\bar{{\rm u}}\bar{{\rm d}}$ dimeson
in LHC, and in the final Section we call for new ideas how to
detect it.

\section{BINDING ENERGIES}

\subsection{With Two-Body Forces}

We estimate the binding energy of the 
${\rm b}{\rm b}\bar{{\rm u}}\bar{{\rm d}}$ dimeson 
assuming the $V_{{\rm QQ}}={1\over2}V_{{\rm Q}\bar{{\rm Q}}}$ rule
and no three-body forces. If this prediction is falsified by
the experiment a revision of the two-body QQ interaction and/or
introduction of many-quark forces will be needed.

For the sake of clarity we present here a simplified
derivation which makes three further assumptions: (i) that the 
spin-dependent interactions between heavy quarks are not important, 
(ii) that the configuration  with both b quarks bound in a compact 
diquark with  spin 1 and antitriplet colour dominates, and
(iii) that the two light antiquarks in the dimeson
behave equally as in a $\Lambda_{{\rm b}}$ baryon. 
We have, however, performed also a careful four-body calculation
without such unnecessary assumptions \cite {JR} and the results
were practically the same.

We want the result to be as little model dependent as possible.
Therefore we consider any model which reproduces exactly the
masses of $\Lambda_{{\rm b}}$ baryon and several mesons. Then we can
consider Nature as an "analogue computer" and express
the mass of the dimeson in terms of those experimental masses.
Now we compare the following hadrons

\begin{eqnarray*}
M_{{\rm bb}\bar{{\rm u}}\bar{{\rm d}}} &=& 
   2M_{{\rm b}} + M_{{\rm u}} +M_{{\rm d}} + E_{{\rm b}{\rm b}} 
  + E_{\bar{{\rm u}}\bar{{\rm d}} [{\rm bb}]} \\
M_\Upsilon  &=& 2M_{{\rm b}}  + E_{{\rm b}\bar{{\rm b}}} \\
M_{\Lambda_{{\rm b}}} &=& M_{{\rm b}} + M_{{\rm u}} +M_{{\rm d}}
  + E_{\bar{{\rm u}}\bar{{\rm d}} \bar{\rm b}}
\end{eqnarray*}
where $E_{\bar{{\rm u}}\bar{{\rm d}} [{\rm bb}]}\approx 
E_{\bar{{\rm u}}\bar{{\rm d}} \bar{\rm b}}$ is the
potential plus kinetic energy contribution of the two light quarks 
in the field of a heavy diquark or quark, respectively,  
and it cancels in the difference. This would be exactly true
in the limit where the mass of the b quark goes to infinity and the
heavy diquark is point-like so that we 
can neglect the size of the heavy diquark in the dimeson.

We can estimate the diquark binding energy using the following trick.
The binding energy $E_{{\rm meson}}$ of a quark and antiquark 
in the meson is a function of the reduced mass only (neglecting spin forces):
$$
\Bigl[\frac{p^2}{M_{{\rm Q}}}+V_{{\rm Q}\bar{{\rm Q}}}\Bigr]\psi
   = E_{{\rm meson}}(M_{{\rm Q}}) \psi
$$

For a diquark the Schr\"odinger equation is similar as for a meson,
but with twice weaker interaction.
To get the similarity, we mimic the kinetic energy with twice
smaller reduced mass. 
$$
\Bigl[\frac{p^2}{M_{\rm Q}}+V_{{\rm Q}{\rm Q}}\Bigr]\psi=\frac{1}{2}\Bigl[
  \frac{p^2}{M_{\rm Q}/2}+V_{{\rm Q}\bar{{\rm Q}}} \Bigr]\psi
  =\frac{1}{2}E_{{\rm meson}}(M_{\rm Q}/2)\psi.
$$
We obtain $E_{{\rm meson}}(M_{\rm b}/2)$ by plotting the binding energies
of mesons
as a function of the reduced mass  \cite{JR}. We do that for
different choices of constituent quark masses in order to estimate the
uncertainty due to quark masses. The plot is very smooth
and we estimate the binding energy of the heavy  bb diquark 
$E_{{\rm diquark}}(M_{\rm b}) = \frac{1}{2} E_{{\rm meson}}(M_{\rm b}/2)=
-390\pm 15$ MeV, whereas for bottomium
$\frac{1}{2} E_{{\rm meson}}(M_{\rm b})=-560\pm 15$ MeV.

This gives us the phenomenological estimate for the binding
energy of the dimeson with respect to the BB$^*$ threshold

\begin{eqnarray*}
\Delta E_{{\rm bb}\bar{{\rm u}}\bar{{\rm d}}}
   &=&M_{\Lambda_{{\rm b}}}+\big[M_{\Upsilon}-
   E_{{\rm meson}}(M_{{\rm b}})+E_{{\rm meson}}(M_{{\rm b}}/2)\big]/2
   -M_{{\rm B}}-M_{{\rm B^*}}\\
   &=& -130\pm 20\, {\rm MeV}.
\end{eqnarray*}
This agrees well with our detailed calculation \cite{JR} and
with some previous four body calculations \cite{SB} \cite{BS}
in the constituent quark model.

An analogous calculation using $\Lambda_{{\rm c}}$ and ${\rm J}/\psi$
instead of $\Lambda_{{\rm b}}$ and $\Upsilon$ gives for the
mass difference between the ${\rm cc}\bar{{\rm u}}\bar{{\rm d}}$
dimeson and the DD$^*$ threshold a value of +97 MeV which means
a prediction, that this dimeson is definitely unbound.

\subsection{With Two- and Three-Body Forces}

Nothing is definite. If a future experiment finds the bb dimeson unbound
or the cc dimeson bound
we shall have to revise our general ideas about
the effective quark-quark interaction, and/or introduce
many-quark forces.

The recent CDF experiment on double-heavy baryons in Fermilab 
\cite{Selex} has caused some confusion.
The ccd candidate at 3520 MeV is somewhat low but still manageable
with two-body interactions. One would need smaller constituent quark
masses than \cite{SB}. Why? Smaller quark masses need less negative
(kinetic + potential) binding energies to fit the heavy mesons. 
\,\,Therefore, going from ${\rm Q}\bar{{\rm Q}}$ to the QQ diquark
using the $V_{{\rm QQ}}={1\over2}V_{{\rm Q}\bar{{\rm Q}}}$ rule
one loses less binding and the ccd baryon becomes better bound.
(Note that a phenomenological estimate similar to the one for the 
dimesons would give 3537 to 3560 MeV).
The ccd candidate at 3783 seems, however, too light for an excited 
state and too heavy for the ground state.

On the other hand, the ccu baryon candidate at 3460 MeV 
is not believable due to its 60 MeV isospin splitting.
If, however, correct, it would require help from many-body
forces or some other mechanisms which would lead to an
almost bound cc dimeson (our phenomenological estimate
can then be written as
$\Delta E_{{\rm cc}\bar{{\rm u}}\bar{{\rm d}}}
   =M_{\Xi_{{\rm cc}}} + M_{\Lambda_{{\rm c}}}
    -M_{{\rm D}}-M_{{\rm D}}-M_{{\rm D^*}}
    = +3\, {\rm MeV}$).

Some weak three-body terms have been introduced in ref.\cite{SB}
in order to improve (lower) the absolute position of the
baryon spectrum if the meson spectrum is fitted.
We are exploring which colour structures could be used
and whether the parametrisation could be stretched  
so as to bind cc dimeson or unbind the bb dimeson 
without spoiling the fit to mesons and baryons.

\section{A MODEL FOR THE SYNTHESIS OF THE HEAVY DIMESON}

The ${\rm bb\bar{u}\bar{d}}$ dimesons have not been seen in the present
machines. Anyway, estimates of the production and detection rate 
are very pessimistic and have for this reason not been published.
Therefore we encourage to look for them at LHC.
For the synthesis  we propose a three-step model \cite{RJTF}\cite{JRTF}. 

{\em (i) First, two b-quarks are formed in the proton-proton collision
by a double gluon-gluon fusion:}
${\rm (g+g) + (g+g)}\to({\rm b} + \bar{{\rm b}})+
({\rm b} + \bar{{\rm b}})$. This was shown to be the dominant
production mechanism \cite{FT}. 
One might wonder why we need a TeV machine to produce GeV particles.
The answer is simple. The two colliding protons can be considered as
two packages of virtual gluons whose number is huge for low Bjorken-$x$.
Only the number of gluons with $x < 0.001$ might be sufficient
to make dimesons detectable.

The forward detector LHCb will cover the pseudorapidity region 
$1.8<\eta<4.9$ and will detect the B and 
$\bar{{\rm B}}$ hadrons in the low $p_{{\rm T}}$ region. 
We are interested in double-b production in which the two b-quarks 
are close enough in phase space to synthesize a diquark.
By requiring that the two b are produced with 
$|p_1(j)-p_2(j)|<\Delta, \; j=x,y,z$, we get
the cross section $\sigma\approx 0.4 (\Delta/{\rm GeV})^3$ nb which is
approximately proportional to the momentum volume up to 2 GeV:
$ d \sigma / d^3 p \approx 0.4\, {\rm nb/GeV}^3.$
At the expected luminosity $L$=0.1 events/(second nb) this corresponds to
144 interesting bb pairs per hour per GeV$^3$.

{\em (ii) In the second step, the two {\rm b} quarks join into a diquark.}
We assume simultaneous production of two independent b quarks
with momenta $\vec{p}_1, \vec{p}_2$. Since they appear wherever within 
the nucleon volume, we modulate their wavefunctions with a Gaussian profile  
with the ``oscillator parameter'' $B=\sqrt{2/3}\sqrt{<r^2>}=0.69$ fm
corresponding to the nucleon rms radius
\begin{eqnarray*}
  {\cal N}_B &\exp& (-\vec{r}_1^{\,2}/2B^2+{\rm i}\, \vec{p}_1 \vec{r}_1)
  {\cal N}_B \exp (-\vec{r}_2^{\,2}/2B^2+{\rm i}\, \vec{p}_2 \vec{r}_2) \\
  \equiv
  {\cal N}_{B/\sqrt{2}} &\exp& (-\vec{R}^2/2(B/\sqrt{2})^2+{\rm i}\, 
    \vec{P} \vec{R})
  {\cal N}_{B\sqrt{2}} \exp (-\vec{r}^{\,2}/2(B\sqrt{2})^2+{\rm i}\, 
    \vec{p} \vec{r})
\end{eqnarray*}
where the normalization factor ${\cal N}_\beta=\pi^{-3/4}\beta^{-3/2}.$

We make an impulse approximation that this two-quark 
state is instantaneously transformed in any of the eigenstates of the
two-quark Hamiltonian. Then the amplitude of the 
diquark formation $M$ is equal to the
overlap  between the two free quarks and the diquark 
with the same centre-of-mass motion. By approximating 
the diquark wavefunction  with a Gaussian with the
oscillator parameter $\beta=0.23$ fm we get 
\begin{eqnarray*}
M(p) &=& \int d^3 r \> 
  {\cal N}_{B\sqrt{2}} \exp (-\vec{r}^{\,2}/2(B\sqrt{2})^2-{\rm i}\, 
     \vec{p} \vec{r})
  {\cal N}_\beta \exp (-\vec{r}^{\,2}/2\beta^2) \\
  &=&  \sqrt{\frac{2\sqrt{2} B\beta}{2B^2+\beta^2}}^3\>
       \exp{[-(p^2/2)(2B^2\beta^2/(2B^2+\beta^2))]}
\end{eqnarray*}

For the production cross section we take into account that
$\beta<<B$ and that
$ d \sigma / d^3 p$ is practically constant and can be taken 
out of the integral
$$
\sigma = \int d^3 p\,\frac{d \sigma}{d^3 p} M^2(p) 
   \approx \frac{d \sigma}{d^3 p} \, 
          \left(\frac{\sqrt{2\pi}\,\hbar}{B}\right)^3 = 0.15 {\rm nb}
$$
which corresponds to $L\sigma=54$ diquarks/hour.

{\em (iii) In the third step, the diquark gets dressed}.
It either acquires one light quark to become 
the doubly-heavy baryon bbu, bbd or bbs, or
two light antiquarks to become a dimeson.

We estimate the probabilities of dressing $f_{{\rm dress}}$ using the 
analogy with dressing a single quark (``fragmentation of a quark into
hadrons''). We make use of experimental data obtained 
at Fermilab and at LEP experiments \cite{Fermi}: 
$$\bar{{\rm b}}\to \bar{{\rm b}}{\rm u},\,\bar{{\rm b}}{\rm d},\,
   \bar{{\rm b}}{\rm s},\,\bar{{\rm b}}\bar{{\rm u}}\bar{{\rm d}}
   = 0.37\pm0.02,\;0.37\pm0.02,\; 0.16\pm0.03,\; 0.09\pm0.03.$$
Since a heavy diquark acts similarly as a heavy quark,
we expect similar branching ratios:
$${\rm bb} \to {\rm bbu,\; bbd,\; bbs,\; bb}\bar{{\rm u}}\bar{{\rm d}} 
\approx 0.37,\; 0.37,\; 0.16,\; 0.09.$$

This yields the production rate of the dimeson
$L\sigma f_{{\rm dress}}\sim 5 - 6$ events/hour.

\section{DISCUSSION ABOUT THE DETECTION OF DIMESONS}

We expect that the  ${\rm bb\bar{u}\bar{d}}$ dimeson will be
stable against strong and electromagnetic decay and will decay only weakly,
with a lifetime of about 1 ps (corresponding to the width of 1 meV).
The main channel would be the independent decay of each b quark 
into c quark. This essentially means the independent decay of each
B meson in the dimeson, for example ${\rm B}\to{\rm D+anything}$.
However, such a channel will be difficult to
distinguish from the decay of two unbound B mesons which will be 
the main background contribution. There are no good two-body decay channels
of B mesons to allow the reconstruction of the total
energy of the dimeson; moreover, each separate 
exclusive decay channel has a low branching ratio of up to a few percent.

Much more characteristic would be the simple two-body decay
channel   ${\rm bb\bar{u}\bar{d}}\to\Upsilon + \pi$
with the kinetic energy 876 MeV of both mesons together (in the c.m. system).
Of course there would be a crowd of other $\Upsilon$ mesons, but few at
this energy.
The inspiration comes from the  ${\rm B^0}\to\bar{{\rm B}}^0$ oscillation 
which unfortunately is not feasible for bound $B$ mesons because the 
BB and B$\bar{{\rm B}}$ states are not degenerate.
The weak transition ${\rm b}\bar{{\rm u}}\to{\rm u}\bar{{\rm b}}$
is negligible because of the low CKM amplitudes. 

Some hope is offered by the angular correlation of the b and c quarks
after the decay of the first b$\to$c, from which the original 
b-b correlation could be deduced. If a dimeson were formed, 
the correlation should be isotropic in the c.m. system of the dimeson 
since the two heavy quarks are expected
to be in the relative 1s state. On the other hand, 
two independent b-quarks would tend to move more in the same direction.

We call for new ideas for the detection of doubly b dimesons!

\end{document}